\documentclass{amsart} 
\usepackage{graphicx}
\usepackage{amssymb, amsmath, sidecap}
\usepackage{amsfonts}
\usepackage{amssymb}
\usepackage{float}

\newtheorem{theorem}{Theorem}

\newtheorem{definition}[theorem]{Definition}

\newcommand{\C}{\mathbb C}
\newcommand{\N}{\mathbb N}

\newcommand{\R}{\mathbb R}

\def\la{\label}
\def\bt{\begin{thm}}
\def\et{\end{thm}}

\def\bl{\begin{lem}}
\def\el{\end{lem}}

\def\bd{\begin{defi}}
\def\ed{\end{defi}}

\def\bc{\begin{cor}}
\def\ec{\end{cor}}

\def\bp{\begin{proof}}
\def\ep{\end{proof}}

\def\br{\begin{rem}}
\def\er{\end{rem}}

\newtheorem{thm}{Theorem}[section]
\newtheorem{lem}{Lemma}[section]
\newtheorem{defi}{Definition}[section]

\newtheorem{rem}{Remark}[section]
\newtheorem{cor}{Corollary}[section]

\numberwithin{equation}{section}
\numberwithin{theorem}{section}
\numberwithin{example}{section}

\numberwithin{figure}{section}

\begin{document}
\title{Dynamic Model and Phase Transitions for Liquid Helium}
\author[Ma]{Tian Ma}
\address[TM]{Department of Mathematics, Sichuan University,
Chengdu, P. R. China}

\author[Wang]{Shouhong Wang}
\address[SW]{Department of Mathematics,
Indiana University, Bloomington, IN 47405}
\email{showang@indiana.edu, http://www.indiana.edu/~fluid}

\thanks{The work was supported in part by the
Office of Naval Research and by the National Science Foundation.}


\begin{abstract}
This article presents a phenomenological dynamic phase transition theory -- modeling and analysis -- for superfluids. As we know, although the time-dependent Ginzburg-Landau model has been successfully used in superconductivity, and the classical Ginzburg-Landau free energy 
is still poorly applicable to liquid helium in a quantitative sense. 
The study in this article is based on 1) a new dynamic classification scheme of phase transitions, 2)  new time-dependent Ginzburg-Landau models  for general equilibrium transitions, and 3) the general dynamic transition theory.  
The results in this article predict the existence of a unstable region $H$, where both solid and liquid He II states appear randomly depending on fluctuations and the existence of a switch point $M$ on the $\lambda$ curve, where the transitions changes types.\end{abstract}
\maketitle

\section{Introduction}
\label{sc1}
Superfluidity is a phase of matter  in which "unusual" effects are observed when liquids, typically of helium-4 or helium-3, overcome friction by surface interaction when at a stage, known as the "lambda point" for helium-4, at which the liquid's viscosity becomes zero. Also known as a major facet in the study of quantum hydrodynamics, it was discovered by Pyotr Leonidovich Kapitsa, John F. Allen, and Don Misener in 1937 and has been described through phenomenological and microscopic theories. 

Atoms helium have two stable isotopes $^4$He and $^3$He. $^4$He 
consists of two electrons, two protons and two neutrons, which are
six fermions. Therefore, $^4$He has an integral spin and obey the
Bose-Einstein statistics. Liquid $^4$He, called the Bose liquid, displays a
direct transition from the normal liquid state (liquid He I) to
the superfluid state (liquid He II) at temperature $T=2.19K$,
which can be considered as the condensation of particles at simple
quantum state. 

The main objectives  of this article are  1) to establish a time-dependent  
Ginzburg-Landau model for liquid $^4$He, and 2) to study its dynamic phase transitions.  
Hereafter, we shall present briefly the main ingredients of the study presented in this article. The ideas and method in this article can be used  to study $^3$He and its mixture  with $^4$He; this study will be reported elsewhere. 

First, in the late 1930s, Ginzburg-Landau proposed a mean field theory of continuous phase transitions.  With the successful application of the Ginzburg-Landau theory to  superconductivity, it is nature to transfer something similar to the superfluidity case, as the superfluid transitions in liquid $^3$He  and  $^4$He   are of similar quantum origin as superconductivity. 
Unfortunately, we know that the classical Ginzburg-Landau free energy 
is poorly applicable to $^4$He in a quantitative sense, as described in by Ginzburg in \cite{ginzburg}.

The starting point of the  modeling used in this article is to introduce a general principle, based on the le Ch\^atelier principle and some general characteristics of pseudo-gradient flow systems. This general principle leads  to a unified approach to derive Ginzburg-Landau type of time-dependent models for equilibrium phase transitions. 
With this general principle in our disposal, we derive some dynamic models for superfluid transitions. 

In addition, an important difference for the new dynamic models introduced here in this article from the classical ones
 is based on the separation of superfluid and normal fluid densities as described in (\ref{8.199}) and their interactions in the Ginzburg-Landau free energies.

Second, 
classically, phase transitions are classified by  the Ehrenfest classification scheme,  based on  the lowest derivative of the free energy that is discontinuous at the transition. 
The superfluid  phase transitions has been regarded as  continuous phase transitions, also called second-order phase transitions. However, many important issues are still still not clear; see among many others  
Ginzburg \cite{ginzburg}, Reichl \cite{reichl}  and  Onuki \cite{onuki}. 

One important new ingredient for the analysis is  a new dynamic transition theory developed recently by the authors \cite{chinese-book, b-book}. With this theory, we derive a  new dynamic phase transition classification scheme, which   classifies phase transitions into three categories:  Type-I, Type-II and Type-III, corresponding respectively to the continuous, the jump and mixed transitions in the dynamic transition theory. 

The results in this article lead to two physical predictions: 1) the existence of the unstable region $H$, where both solid and liquid He II states appear randomly depending on fluctuations, and 2) the existence of a switch point $M$, where the transitions,  between superfluid state (liquid He II) and the normal fluid state (liquid He I), changes from first order (Type II with the dynamic classification scheme)  
to second order (Type-I). Of course, these predictions need  to be verified by experiments, and it is hoped that the new model, the ideas  and methods introduced in this article  will lead to some improved understanding of superfluidity.

This article is organized as follows.
First, the new dynamic transition theory is recapitulated in Section 2, and a  dynamic phase transition model for $^4$He is introduced in discussed in Section 3. The dynamic transitions of the the model is analyzed in Section 4. Section 5 gives some physical conclusions and predictions.  

\section{General Principles of Phase Transition Dynamics}
In this section,  we introduce a new phase dynamic transition classification scheme 
to classify phase transitions into three categories: Type-I, Type-II and Type-III, corresponding mathematically  continuous,  jump mixed transitions, respectively.

\subsection{Dynamic transition theory}
In sciences, nonlinear dissipative systems are generally governed
by differential equations, which can be expressed in the following
abstract form 
Let $X$  and $ X_1$ be two Banach spaces,   and $X_1\subset X$ a compact and
dense inclusion. In this chapter, we always consider the following
nonlinear evolution equations
\begin{equation}
\frac{du}{dt}=L_{\lambda}u+G(u,\lambda),\qquad u(0)=\varphi ,
\label{5.1}
\end{equation}
where $u:[0,\infty )\rightarrow X$ is unknown function,  and 
$\lambda\in \R^1$  is the system parameter.

Assume that $L_{\lambda}:X_1\rightarrow X$ is a parameterized
linear completely continuous field depending contiguously on
$\lambda\in \R^1$, which satisfies
\begin{equation}
\left. 
\begin{aligned} 
&L_{\lambda}=-A+B_{\lambda}   && \text{a sectorial operator},\\
&A:X_1\rightarrow X   && \text{a linear homeomorphism},\\
&B_{\lambda}:X_1\rightarrow X&&  \text{a linear compact  operator}.
\end{aligned}
\right.\label{5.2}
\end{equation}
In this case, we can define the fractional order spaces
$X_{\sigma}$ for $\sigma\in \R^1$. Then we also assume that
$G(\cdot ,\lambda ):X_{\alpha}\rightarrow X$ is $C^r(r\geq 1)$
bounded mapping for some $0\leq\alpha <1$, depending continuously
on $\lambda\in \R^1$, and
\begin{equation}
G(u,\lambda )=o(\|u\|_{X_{\alpha}}) \qquad  \forall\lambda\in
\R^1.\label{5.3}
\end{equation}

Hereafter we always assume the conditions (\ref{5.2}) and
(\ref{5.3}), which represent that the system (\ref{5.1}) has
a dissipative structure.

A state of the system (\ref{5.1}) at $\lambda$ is usually referred to as a compact invariant set $\Sigma_{\lambda}$. In many applications, 
$\Sigma_{\lambda}$ is a singular point or a periodic orbit. A state $\Sigma_{\lambda}$
of (\ref{5.1}) is stable if $\Sigma_{\lambda}$ is an attractor;
otherwise $\Sigma_{\lambda}$ is called unstable.

\begin{defi}
\label{d7.1}
We say that the system (\ref{5.1}) has a
phase transition from a state $\Sigma_{\lambda}$ at $\lambda
=\lambda_0$ if $\Sigma_{\lambda}$ is stable on $\lambda <\lambda_0$
(or on $\lambda >\lambda_0$) and is unstable on $\lambda
>\lambda_0$ (or on $\lambda <\lambda_0$).  The critical parameter $\lambda_0$ is
called a critical point.
 In other words, the
phase transition corresponds to an exchange of stable states.
\end{defi}

Obviously, the attractor bifurcation of (\ref{5.1}) is a type of
transition. However,  bifurcation and
transition are two different, but related concepts. 

Let $\{\beta_j(\lambda )\in \C\ \   |\ \ j \in \N\}$  be the eigenvalues (counting multiplicity) of $L_{\lambda}$, and  assume that
\begin{align}
&  \text{Re}\ \beta_i(\lambda )
\left\{ 
 \begin{aligned} 
 &  <0 &&    \text{ if } \lambda  <\lambda_0,\\
& =0 &&      \text{ if } \lambda =\lambda_0,\\
& >0&&     \text{ if } \lambda >\lambda_0,
\end{aligned}
\right.   &&  \forall 1\leq i\leq m,  \label{5.4}\\
&\text{Re}\ \beta_j(\lambda_0)<0 &&  \forall j\geq
m+1.\label{5.5}
\end{align}

The following theorem is a basic principle of transitions from
equilibrium states, which provides sufficient conditions and a basic
classification for transitions of nonlinear dissipative systems.
This theorem is a direct consequence of the center manifold
theorems and the stable manifold theorems; we omit the proof.

\bt\la{t5.1}
 Let the conditions (\ref{5.4}) and
(\ref{5.5}) hold true. Then, the system (\ref{5.1}) must have a
transition from $(u,\lambda )=(0,\lambda_0)$, and there is a
neighborhood $U\subset X$ of $u=0$ such that the transition is one
of the following three types:

\begin{itemize}
\item[(1)] {\sc Continuous Transition}: 
there exists an open and dense set
$\widetilde{U}_{\lambda}\subset U$ such that for any
$\varphi\in\widetilde{U}_{\lambda}$,  the solution
$u_{\lambda}(t,\varphi )$ of (\ref{5.1}) satisfies
$$\lim\limits_{\lambda\rightarrow\lambda_0}\limsup_{t\rightarrow\infty}\|u_{\lambda}(t,\varphi
)\|_X=0.$$ 

\item[(2)] {\sc Jump Transition}: 
for any $\lambda_0<\lambda <\lambda_0+\varepsilon$ with some $\varepsilon >0$, there is an open
and dense set $U_{\lambda}\subset U$ such that 
for any $\varphi\in U_{\lambda}$, 
$$\limsup_{t\rightarrow\infty}\|u_{\lambda}(t,\varphi
)\|_X\geq\delta >0 \quad \text{ for some $\delta >0$ is independent of $\lambda$}.$$ 

\item[(3)] {\sc Mixed Transition}: 
for any $\lambda_0<\lambda <\lambda_0+\varepsilon$  with some $\varepsilon >0$, 
$U$ can be decomposed into two open sets
$U^{\lambda}_1$ and $U^{\lambda}_2$  ($U^{\lambda}_i$ not necessarily
connected):
$\bar{U}=\bar{U}^{\lambda}_1+\bar{U}^{\lambda}_2$, $U^{\lambda}_1\cap U^{\lambda}_2=\emptyset$,  such that
\begin{align*}
&\lim\limits_{\lambda\rightarrow\lambda_0}\limsup_{t\rightarrow\infty}\|u(t,\varphi
)\|_X=0   &&   \forall\varphi\in U^{\lambda}_1,\\
& \limsup_{t\rightarrow\infty}\|u(t,\varphi
)\|_X\geq\delta >0 && \forall\varphi\in U^{\lambda}_2.
\end{align*}
\end{itemize}
\et

With this theorem in our disposal, we are in position to give a new dynamic classification scheme for dynamic phase transitions.

\begin{definition}[Dynamic Classification of Phase Transition]
The phase transitions for  (\ref{5.1}) at $\lambda =\lambda_0$ is classified using  their  dynamic properties: continuous, jump, and mixed as given in Theorem~\ref{t5.1}, which are called Type-I, Type-II and Type-III respectively.
\end{definition}

An important aspect of the  transition theory is to determine which 
of the three types of transitions given by Theorem \ref{t5.1} occurs in
a specific  problem. A corresponding dynamic transition has been developed recently by the authors for this purpose; see \cite{chinese-book}. We refer interested readers to these references for details of the theory. Hereafter we recall a few related theorems in this theory used in this article.

\medskip

{\sc First}, we remark here that one crucial ingredient for applications of this general dynamic transition theory is  
the reduction of (\ref{5.1}) to the center manifold function.
In fact, by  this reduction, we know that the type of transitions for (\ref{5.1}) at
$(0,\lambda_0)$ is completely dictated  by its reduction equation
near $\lambda =\lambda_0$:
\begin{equation}
\frac{dx}{dt}=J_{\lambda}x+g(x,\lambda ) \qquad  \text{ for } x \in \R^m,\label{5.7}
\end{equation}
where  $g(x,\lambda )=(g_1(x,\lambda ),\cdots
,g_m(x,\lambda ))$,   and 
\begin{equation}
g_j(x,\lambda )= <G(\sum^m_{i=1}x_ie_i + \Phi (x,\lambda ),\lambda), e^*_j> 
\quad   \forall 1\leq j\leq m.
\label{5.8}
\end{equation}
Here $e_j$ and $e^*_j$  $(1\leq j\leq m)$ are the eigenvectors of
$L_{\lambda}$ and $L^*_{\lambda}$ respectively corresponding to the
eigenvalues $\beta_j(\lambda )$ as in (\ref{5.4}),  $J_{\lambda}$ is the $m\times m$ order Jordan matrix
corresponding to the eigenvalues given by (\ref{5.4}),    and $\Phi
(x,\lambda )$ is the center manifold function of (\ref{5.1}) near
$\lambda_0$. 

\medskip

{\sc Second},  let (\ref{5.1}) be a gradient-type equation. Under the conditions (\ref{5.4}) and (\ref{5.5}), 
in a
neighborhood $U\subset X$ of $u=0$, the center manifold $M^c$ in
$U$ at $\lambda =\lambda_0$ consists of  three subsets
$$M^c=W^u+W^s+D,$$
where $W^s$ is the stable set, $W^u$ is the unstable set, and $D$ is  the
hyperbolic set of (\ref{5.7}). Then we have the following
theorem.

\bt\la{t5.3} 
Let (\ref{5.1}) be a gradient-type equation,
and the conditions (\ref{5.4}) and (\ref{5.5}) hold true. If $u=0$
is an isolated singular point of (\ref{5.1}) at $\lambda
=\lambda_0$, then we have the following assertions:

\begin{itemize}

\item[(1)] The transition of (\ref{5.1}) at $(u,\lambda)=(0,\lambda_0)$ is continuous if and only if $u=0$ is locally asymptotically stable at $\lambda =\lambda_0$, i.e., the center
manifold is stable: $M^c=W^s$. Moreover, (\ref{5.1}) 
bifurcates from $(0,\lambda_0)$ to minimal attractors consisting  of
singular points of (\ref{5.1}). 

\item[(2)] If the stable set $W^s$ of (\ref{5.1}) has no interior points in $M^c$, i.e.,
$M^c=\bar{W}^u+\bar{D}$, then the transition is jump.
\end{itemize}
\et

{\sc Third}, we also a dynamic transition theorem  of (\ref{5.1}) from a simple critical
eigenvalue, which has been used in analyzing PVT systems and the ferromagnetic systems \cite{MW08c, MW08f}. We refer the interested readers to these references for details for this theorem.

\subsection{New Ginzburg-Landau models for equilibrium phase transitions}
\label{s7.2.2}
In this section, we recall a new time-dependent Ginzburg-Landau theory for modeling  equilibrium phase transitions in statistical physics. 

Consider a thermal system with a control  parameter $\lambda$. The classical  
le Ch\^atelier  principle amounts to saying that 
for a stable  equilibrium state of a system $\Sigma$, when the system deviates from
$\Sigma$ by a small perturbation or fluctuation, there will be a
resuming force to restore this system to return to the stable state
$\Sigma$.

By the mathematical characterization of gradient systems and the le Ch\^atelier principle, for a system with
thermodynamic potential ${\mathcal{H}}(u,\lambda )$, the governing
equations are essentially determined by the functional
${\mathcal{H}}(u,\lambda )$.
When the order parameters $(u_1,\cdots,u_m)$ are nonconserved
variables, i.e., the integers
$$\int_{\Omega}u_i(x,t)dx=a_i(t)\neq\text{constant}.$$
then the time-dependent equations are given by
\begin{equation}
\left.
\begin{aligned} 
&\frac{\partial u_i}{\partial
t}=-\beta_i\frac{\delta}{\delta u_i}{\mathcal{H}}(u,\lambda
)+\Phi_i(u,\nabla u,\lambda ) && \text{ for } 1 \le i \le m\\
&\frac{\partial u}{\partial n}|_{\partial\Omega}=0\ \ \ \
(\text{or}\ u|_{\partial\Omega}=0),
\end{aligned}
\right.\label{7.30}
\end{equation}
where $\delta /\delta u_i$ are the variational derivative,
$\beta_i>0$ and $\Phi_i$ satisfy
\begin{equation}
\int_{\Omega}\sum_i\Phi_i\frac{\delta}{\delta
u_i}{\mathcal{H}}(u,\lambda )dx=0.\label{7.31}
\end{equation}
The condition (\ref{7.31})  is  required by
the Le Ch\^atelier principle. In the concrete problem, the terms
$\Phi_i$ can be determined by physical laws and (\ref{7.31}). We remark here that following the le Ch\^atelier principle, one should have an inequality constraint. However   physical systems often obey most simplified rules, as  many existing models for specific problems are consistent with the equality constraint here. This remark applies to the constraint (\ref{7.37}) below as well.

When the order parameters are the number density and the system
has no material exchange with the external, then $u_j$  $(1\leq j\leq
m)$ are conserved, i.e.,
\begin{equation}
\int_{\Omega}u_j(x,t)dx=\text{constant}.\label{7.32}
\end{equation}
This conservation law requires a continuity equation
\begin{equation}
\frac{\partial u_j}{\partial t}=-\nabla\cdot J_j(u,\lambda
),\label{7.33}
\end{equation}
where $J_j(u,\lambda )$ is the flux of component $u_j$. In
addition, $J_j$ satisfy
\begin{equation}
J_j=-k_j\nabla (\mu_j-\sum_{i\neq j}\mu_i),\label{7.34}
\end{equation}
where $\mu_l$ is the chemical potential of component $u_l$, 
\begin{equation}
\mu_j-\sum_{i\neq j}\mu_i=\frac{\delta}{\delta
u_j}{\mathcal{H}}(u,\lambda )-\phi_j(u,\nabla u,\lambda
), \label{7.35}
\end{equation}
and  $\phi_j(u,\lambda )$ is a function depending on the other
components $u_i$ $(i\neq j)$. Thus, from
(\ref{7.33})-(\ref{7.35}) we obtain the dynamical equations as
follows
\begin{equation}
\left.
\begin{aligned} 
&\frac{\partial u_j}{\partial
t}=\beta_j\Delta\left[\frac{\delta}{\delta
u_j}{\mathcal{H}}(u,\lambda )-\phi_j(u,\nabla u,\lambda )\right]  && 
\text{ for } 1 \le j \le m,\\
&\frac{\partial u}{\partial n}|_{\partial\Omega}=0,\ \ \ \
\frac{\partial\Delta u}{\partial n}|_{\partial\Omega}=0
\end{aligned}
\right.\label{7.36}
\end{equation}
where $\beta_j>0$ are constants,  and  $\phi_j$ satisfy
\begin{equation}
\int_{\Omega}\sum_j\Delta\phi_j\cdot\frac{\delta}{\delta
u_j}{\mathcal{H}}(u,\lambda )dx=0.\label{7.37}
\end{equation}

When $m=1$, i.e., the system is a binary system, consisting of
two components $A$ and $B$, then the term $\phi_j=0$. The above model covers the classical Cahn-Hilliard model. It is worth mentioning that for multi-component systems, these $phi_j$ play an important rule in deriving good time-dependent models.

If the order parameters $(u_1,\cdots,u_k)$ are coupled to the
conserved variables $(u_{k+1},\cdots,u_m)$, then the dynamical
equations are
\begin{equation}
\left.
\begin{aligned} 
&\frac{\partial u_i}{\partial t}
   =-\beta_i\frac{\delta}{\delta u_i}{\mathcal{H}}(u,\lambda)+\Phi_i(u,\nabla u,\lambda )
   && \text{ for } 1 \le i \le k,\\
& \frac{\partial u_j}{\partial t}
  =\beta_j\Delta\left[\frac{\delta}{\delta u_j}{\mathcal{H}}(u,\lambda )
    -\phi_j(u,\nabla u,\lambda )\right] && \text{ for } k+1 \le j \le m,\\
&\frac{\partial u_i}{\partial n}|_{\partial\Omega}=0\ \ \ \
(\text{or}\ u_i|_{\partial\Omega}=0) && \text{ for } 1 \le i \le k,\\
&\frac{\partial u_j}{\partial n}|_{\partial\Omega}=0,\ \ \ \
\frac{\partial\Delta u_j}{\partial n}|_{\partial\Omega}=0 && \text{ for } k+1 \le j \le m.
\end{aligned}
\right.\label{7.38}
\end{equation}
Here $\Phi_i$  and $\phi_j$ satisfy (\ref{7.31})   and (\ref{7.37}), respectively.

The model (\ref{7.38}) we derive here  gives a general form of the governing
equations to thermodynamic phase transitions, and will play crucial role in studying the dynamics of equilibrium phase transition in statistic physics.

\section{Dynamic Model for Liquid $^4$He}

$^4$He  was first liquidized at $T=4.215K$ and $p=1\times 10^5$(Pa)
by Karmerlingh Onnes in 1908. In 1938, P.L.Kapitza found that when
the temperature $T$ decreases below $T_C=2.17K$, the liquid $^4$He 
will transit from normal liquid state to superfluid state, in
which the fluid possesses zero viscosity i.e., the viscous
coefficient $\eta =0$. The liquids with $\eta =0$ are called the
superfluids, and the flow without drag is called the
superfluidity. The superfluid transition is called $\lambda$-phase
transition, and its phase diagram is illustrated by Figure \ref{f8.33}.
\begin{SCfigure}[25][t]
  \centering
  \includegraphics[width=0.4\textwidth]{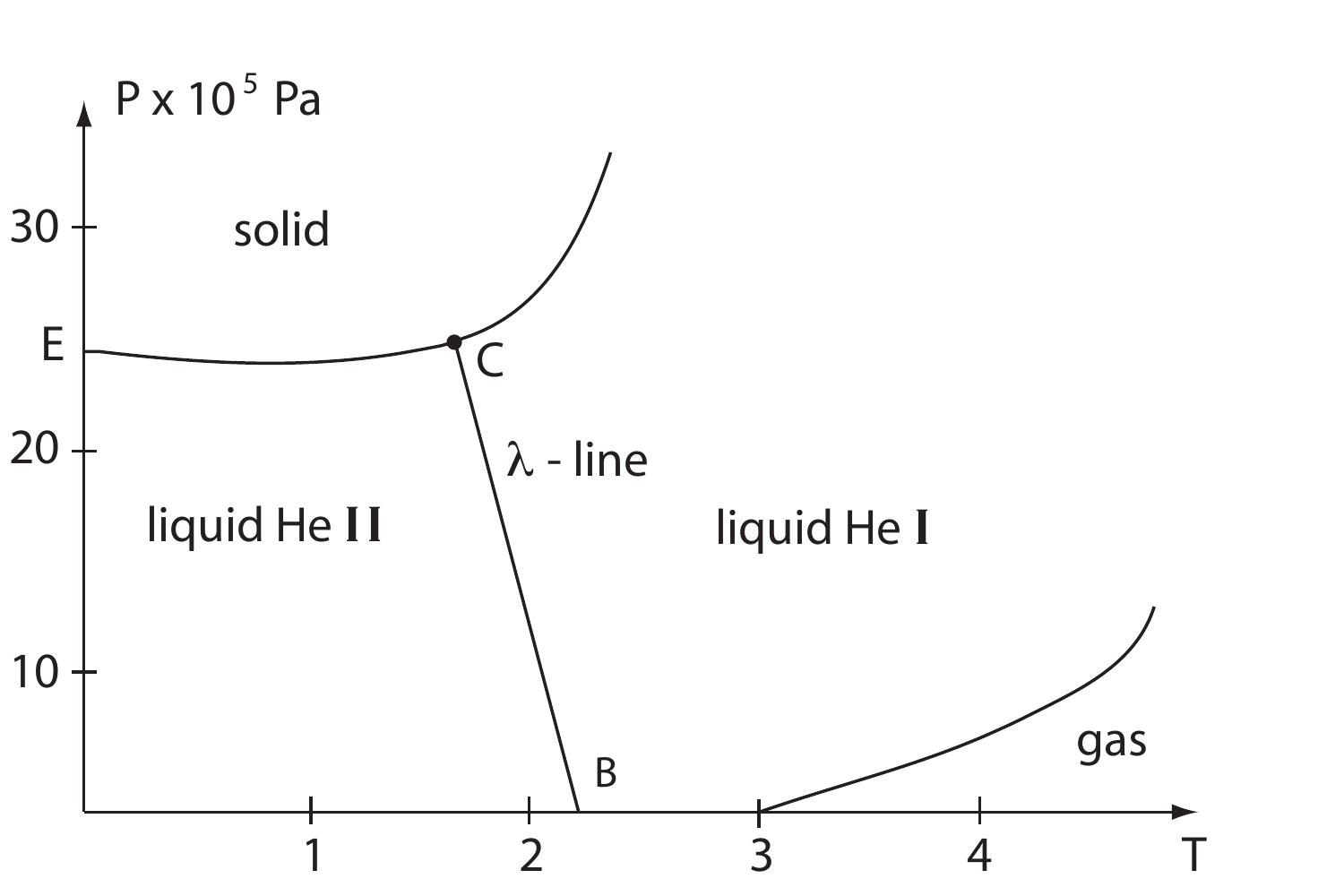}
  \caption{Classical $PT$ phase diagram of $^4$He.}\la{f8.33}
 \end{SCfigure}

\subsection{Ginzburg-Landau free energy}
The order parameter describing superfluidity is characterized by a
non vanishing complex valued function
$\psi:\Omega\rightarrow\mathbb{C}$ as in the
superconductivity, originating from quantum Bose-Einstein
condensation. In the two-fluid hydrodynamic theory of
super-fluidity, the density $\rho$ of $^4$He is given by
\begin{equation}
\rho =\rho_s+\rho_n,\label{8.199}
\end{equation}
where $\rho_s$ is the superfluid density and $\rho_n$ the normal
fluid density. The square $|\psi |^2$ is proportional to $\rho_s$,
and without loss of generality, we take $\psi$ as
\begin{equation}
|\psi |^2=\rho_s.\label{8.200}
\end{equation}

Based on the quantum mechanics, $-ih\nabla\psi$ represents the
momentum associated with the Bose-Einstein condensation. Hence,
the free energy density contains the term
$$\frac{1}{2m}|-ih\nabla\psi |^2=\frac{h^2}{2m}|\nabla\psi |^2,$$
where $h$ is the Planck constant, and $m$ the mass of atom $^4$He .
Meanwhile the superfluid state does not obey the classical
thermodynamic laws, which  the normal liquid state obeys.
Therefore, in the free energy density, $\psi$ satisfies the
Ginzburg-Landau expansion
$$\frac{k_1h^2}{2m}|\nabla\psi |^2+\frac{\gamma_1}{2}|\psi
|^2+\frac{\gamma_2}{4}|\psi |^4,$$ 
and $\rho_n$ has the expansion as in the free energy 
for  $PVT$ systems \cite{MW08c}. For simplicity we
ignore the entropy, and consider the coupling action of $\psi$ and
$\rho_n$, i.e., add the term $\frac{1}{2}\mu\rho_n|\psi |^2$ in the
free energy density. Thus the Ginzburg-Landau free energy for
liquid $^4$He near the superfluid transition is given by
\begin{align}
G(\psi ,\rho_n)=
& 
\int_{\Omega}\Big[\frac{k_1h^2}{2m}|\nabla\psi |^2
    +\frac{\gamma_1}{2}|\psi |^2+\frac{\gamma_2}{4}|\psi |^4 
+\frac{\gamma_3}{2}\rho_n|\psi |^2 \label{8.201}\\
&+\frac{k_2}{2}|\nabla\rho_n|^2+\frac{\mu_1}{2}\rho^2_n
 +\frac{\mu_2}{3}\rho^3_n+\frac{\mu_3}{4}\rho^4_n 
 -p(\rho_n+\frac{\mu_0}{2}\rho^2_n)\Big]dx. \nonumber 
\end{align}

\subsection{Dynamic model governing the superfluidity}

By (\ref{7.30}), we derive from (\ref{8.201}) the following time-dependent 
Ginzburg-Landau equations
governing the superfluidity of liquid $^4$He :
\begin{equation}
\left.
\begin{aligned} &\frac{\partial\psi}{\partial
t}=\frac{k_1h^2}{m}\Delta\psi -\gamma_1\psi -\gamma_2|\psi |^2\psi
-\gamma_3\rho_n\psi ,\\
&\frac{\partial\rho_n}{\partial
t}=k_2\Delta\rho_n-(\mu_1-p\mu_0)\rho_n-\mu_2\rho^2_n-\mu_3\rho^3_n
-\frac{\gamma_3}{2}|\psi
|^2+p.
\end{aligned}
\right.\label{8.202}
\end{equation}

It is known that the following problem has a solution $\rho^0_n\in
H^2(\Omega )\cap H^1_0(\Omega )$ standing for the density of
liquid He I for any $p\in L^2(\Omega )$:
\begin{equation}
\left.
\begin{aligned}
&-k_2\Delta\rho^0_n+(\mu_1-p\mu_0)\rho^0_n+\mu_2(\rho^0_n)^2
+\mu_3(\rho^0_n)^3=p\\
&\left.\frac{\partial\rho^0_n}{\partial
n}\right|_{\partial\Omega}=0.
\end{aligned}
\right.\label{8.203}
\end{equation}

To derive the nondimensional form of  (\ref{8.202}), let
\begin{align*}
& (x, t)=( lx^{\prime}, \tau t^{\prime}),
   &&(\psi, \rho_n) =(\psi_0\psi^{\prime}, \rho_0\rho^{\prime}_n+\rho^0_n), \\
 & \tau =\frac{ml^2}{h^2k_1},&& \mu =\frac{k_2\tau}{l^2}, \\
  &a_1=\gamma_3\rho_0\tau , && a_2=\gamma_2|\psi_0|^2\tau ,\\
 &b_1=\frac{\gamma_3|\psi_0|^2}{2\rho_0}\tau,
          &&b_2=\tau\rho_0(3\rho^0_n\mu_3+\mu_2),\\
  & b_3=\rho^2_0\mu_3\tau , &&\lambda_1=-\tau (\gamma_1+\gamma_3\rho^0_n), \\
  &\lambda_2=\tau (3 (\rho^0_n)^2\mu_3+2\rho^0_n\mu_2+\mu_0p-\mu_1),
\end{align*} 
where $\rho^0_n$ is the solution of (\ref{8.203}).

Thus, suppressing the primes,  the equations (\ref{8.202}) are rewritten as
\begin{equation}
\left.
\begin{aligned} &\frac{\partial\psi}{\partial t}=\Delta\psi
+\lambda_1\psi -a_1\rho_n\psi -a_2|\psi |^2\psi ,\\
&\frac{\partial\rho_n}{\partial
t}=\mu\Delta\rho_n+\lambda_2\rho_n-b_1|\psi
|^2-b_2\rho^2_n-b_3\rho^3_n.
\end{aligned}
\right.\label{8.204}
\end{equation}

The boundary conditions associated with (\ref{8.204}) are
\begin{equation}
\frac{\partial\psi}{\partial n} =0,\ \ \ \
\frac{\partial\rho_n}{\partial
n}=0 \qquad \text{ on } \partial\Omega.\label{8.205}
\end{equation}

When the pressure $p$ is independent of $x\in\Omega$, then the
problem (\ref{8.204}) and (\ref{8.205}) can be approximatively
replaced by the following systems of ordinary differential equations
for  superfluid transitions:
\begin{equation}
\left.
\begin{aligned} &\frac{d\psi}{dt}=\lambda_1\psi
-a_1\rho_n\psi -a_2|\psi |^2\psi ,\\
&\frac{d\rho_n}{dt}=\lambda_2\rho_n-b_1|\psi
|^2-b_2\rho^2_n-b_3\rho^3_n.
\end{aligned}
\right.\label{8.206}
\end{equation}

By multiplying $\psi^*$ to both sides of the first equation of
(\ref{8.206}) and by (\ref{8.200}),  the equations (\ref{8.206})
are reduced to:
\begin{equation}
\left.
\begin{aligned}
&\frac{d\rho_s}{dt}=\lambda_1\rho_s-a_1\rho_n\rho_s-a_2\rho^2_s,\\
&\frac{d\rho_n}{dt}=\lambda_2\rho_n-b_1\rho_s-b_2\rho^2_n-b_3\rho^3_n,\\
&(\rho_s(0),  \rho_n(0)) =( x_0, y_0).
\end{aligned}
\right.\label{8.207}
\end{equation}
for $\rho_s\geq 0$  and   $x_0\geq 0$.

We need to explain the physical properties of the coefficients in
(\ref{8.202}) and (\ref{8.204}). It is known that the coefficients
$\gamma_i$ $(1\leq i\leq 3)$ and $\mu_j$ $(0\leq j\leq 3)$ depend
continuously on the temperature $T$ and the  pressure $p$:
$$
\gamma_i=\gamma_i(T,p),\ \ \ \ \mu_j=\mu_j(T,p)\qquad   \forall  1\leq i\leq 3,\quad  0\leq j\leq 3.
$$ 
From the both mathematical and physical
points of view, the following conditions are required:
\begin{equation}
\gamma_2>0,\quad  \gamma_3>0,\quad  \mu_3>0\qquad  \forall  T,  \rho.\label{8.208}
\end{equation}
In addition, by the Landau mean field theory we have
\begin{align}
& 
\gamma_1\left\{\begin{aligned}
&  >0   &&  \text{  if  }  (T,p)\in A_1,\\
& <0   &&  \text{  if  }    (T,p)\in A_2,
\end{aligned}
\right.\label{8.209} \\
& 
\mu_1\left\{\begin{aligned}
& >0  &&  \text{  if  }(T,p)\in B_1,\\
& <0   &&  \text{  if  }   (T,p)\in B_2,
\end{aligned}
\right.\label{8.210}
\end{align}
where $A_i$, $B_i$ ($i=1,2$)  are connected open sets such that 
$\bar{A}_1+\bar{A}_2=\bar{B}_1+\bar{B}_2=\R^2_+$, 
and $\bar{A}_1\cap\bar{A}_2,
\bar{B}_1\cap\bar{B}_2$ are two simple curves in $\R^2_+$; see Figure
~\ref{f8.34} (a) and (b). In particular, in the PT-plane, 
\begin{equation}
\frac{\partial \gamma_1}{\partial T}>0,\quad \frac{\partial \gamma_1}{\partial p}>0, \quad 
\frac{\partial\mu_1}{\partial T}>0,\quad 
\frac{\partial\mu_1}{\partial p}>0. \label{8.211}
\end{equation}

By (\ref{8.208}) the following nondimensional parameters are
positive
\begin{equation}
a_1>0,\ \ \ \ a_2>0,\ \ \ \ b_1>0,\ \ \ \ b_3>0.\label{8.212}
\end{equation}
By (\ref{8.209})-(\ref{8.211}), the following two critical parameter equations 
\begin{equation}
\left.\begin{aligned}
&
\lambda_1=\lambda_1(T,p)=0,\\
&
\lambda_2=\lambda_2(T,p)=0,
\end{aligned}
\right. \label{8.213}
\end{equation}
give two simple curves $l_1$ and $l_2$ respectively in the $PT$-plane
$\R^2_+$; see Figure~\ref{f8.34} (c) and (d).

The parameters $\mu_2$ and $b_2$ depend on the physical properties
of the atom He, and  satisfy the following relations:
\begin{equation}
\left.
\begin{aligned}
&  b_2(T,p)=\tau\rho_0(\mu_2+3\rho^0_n\mu_3)<0 && \text{ iff }\ 
\rho_{sol}>\rho_l\ \text{at}\ \lambda_2(T,p)=0,\\
& b_2(T,p)=\tau\rho_0(\mu_2+3\rho^0_n\mu_3)>0&& \text{ iff } \ 
\rho_{sol}<\rho_l\
\text{at}\ \lambda_2(T,p)=0,
\end{aligned}
\right.\label{8.214}
\end{equation}
where $\rho_{sol}$ and $\rho_l$ are the densities of solid and
liquid, and $\rho^0_n$ the solution of (\ref{8.203}) representing
the liquid density. These relations in (\ref{8.214}) can be
deduced by the  dynamic transition theorem  of (\ref{5.1}) from a simple critical
eigenvalue, Theorem A.2  in Ma and Wang \cite{MW08f}.

\section{Dynamic phase transition for liquid $^4$He }

In order to illustrate the main ideas, we  discuss only the case where the pressure $p$ is independent of
$x\in\Omega$, i.e., we only consider the equations (\ref{8.206})
and (\ref{8.207}); the general case can be studied in the same fashion and will be reported elsewhere.

\subsection{$PT$-phase diagram}

Based on physical experiments together with
(\ref{8.209})-(\ref{8.211}), the curves of $\gamma_1(T,p)=0$ and
$\mu_1(T,p)=0$ in  the  $PT$-plane are given by Figure \ref{f8.34}(a) and (b)
respectively. By the formulas
\begin{align*}
& \lambda_1(T,p)=-\tau (\gamma_1(T,p)+\gamma_3\rho^0_n),\\
& \lambda_2(T,p)=-\tau
(\mu_1(T,p)-\mu_0p-2\rho^0_n\mu_2-3\rho^{02}_n\mu_3),
\end{align*}
 together
with (\ref{8.208}) and (\ref{8.214}), the curves $l_1$ and $l_2$
given by (\ref{8.213}) in  the  $PT$-plane are illustrated in Figure
\ref{f8.34}(c) and (d).
\begin{figure}[hbt]
  \centering
  \includegraphics[width=0.35\textwidth]{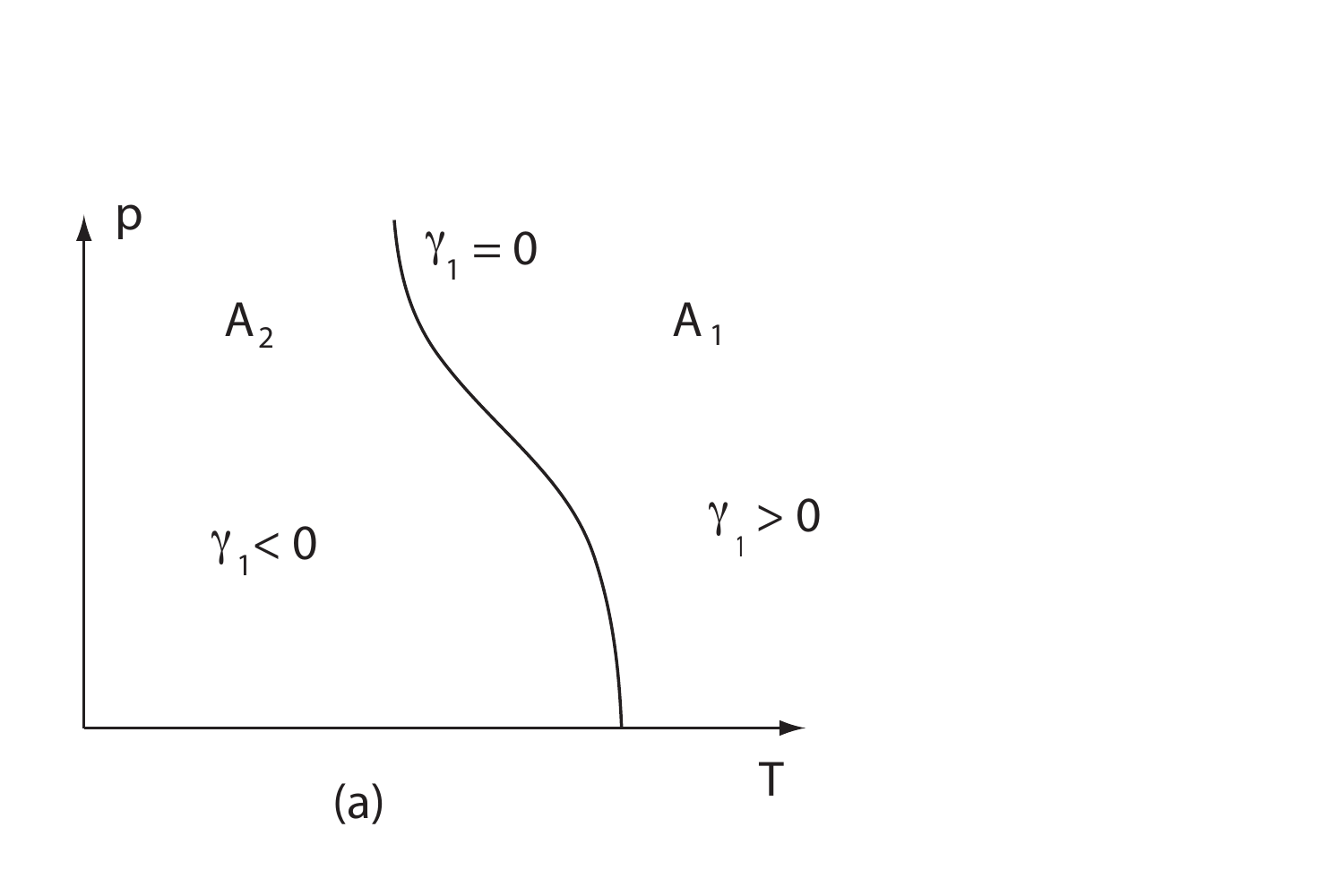}\qquad
   \includegraphics[width=0.38\textwidth]{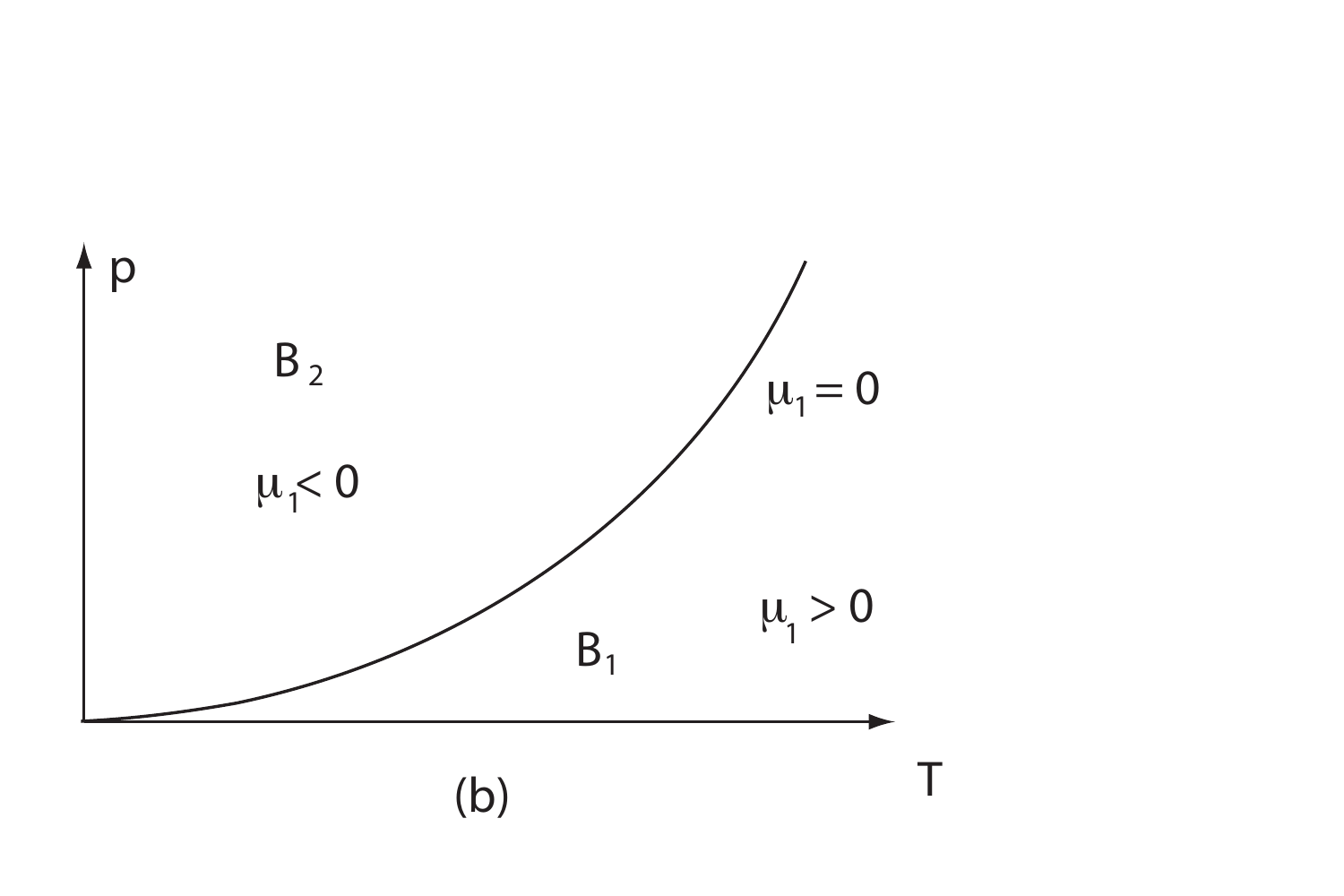}\\
   \includegraphics[width=0.35\textwidth]{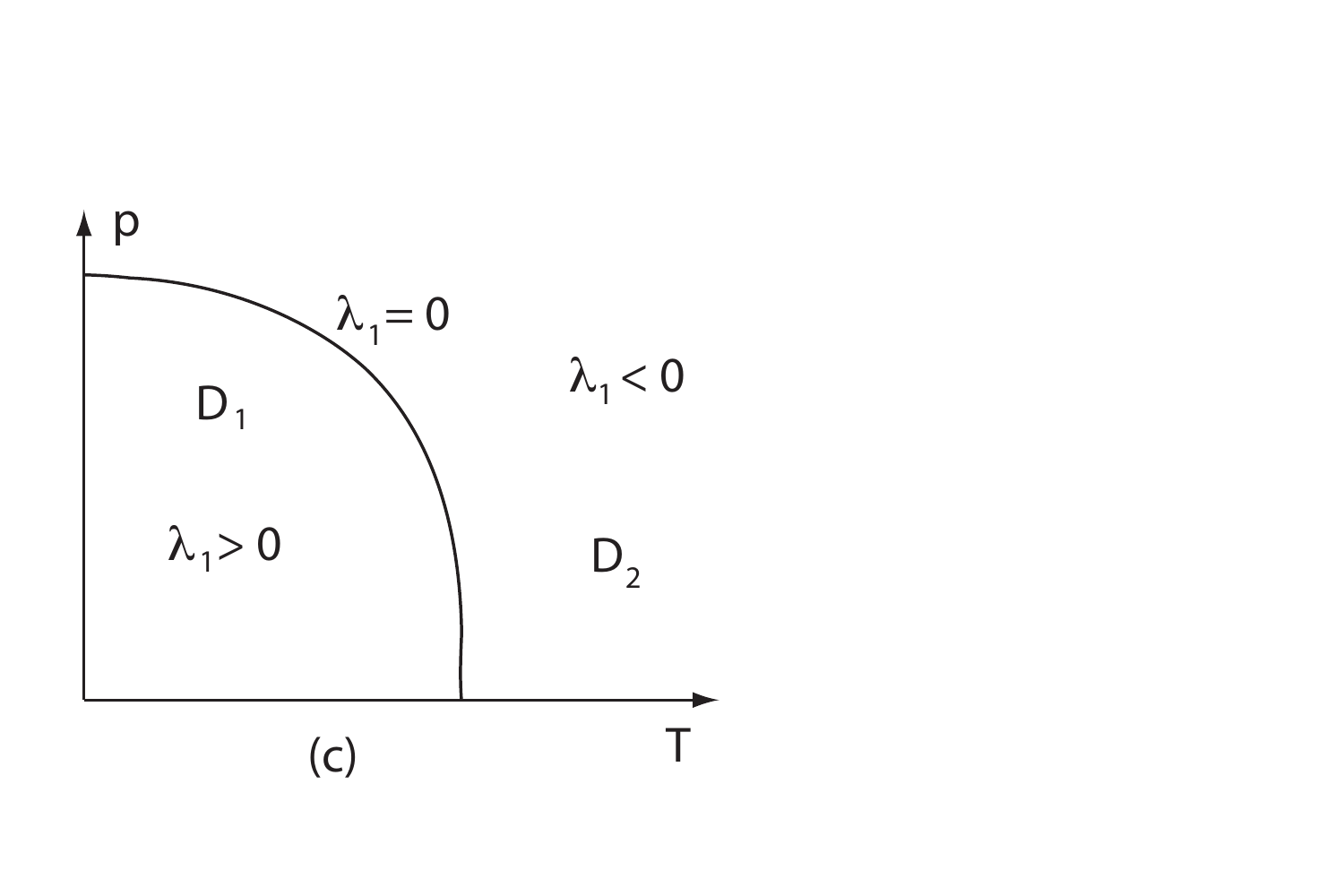}\qquad
   \includegraphics[width=0.35\textwidth]{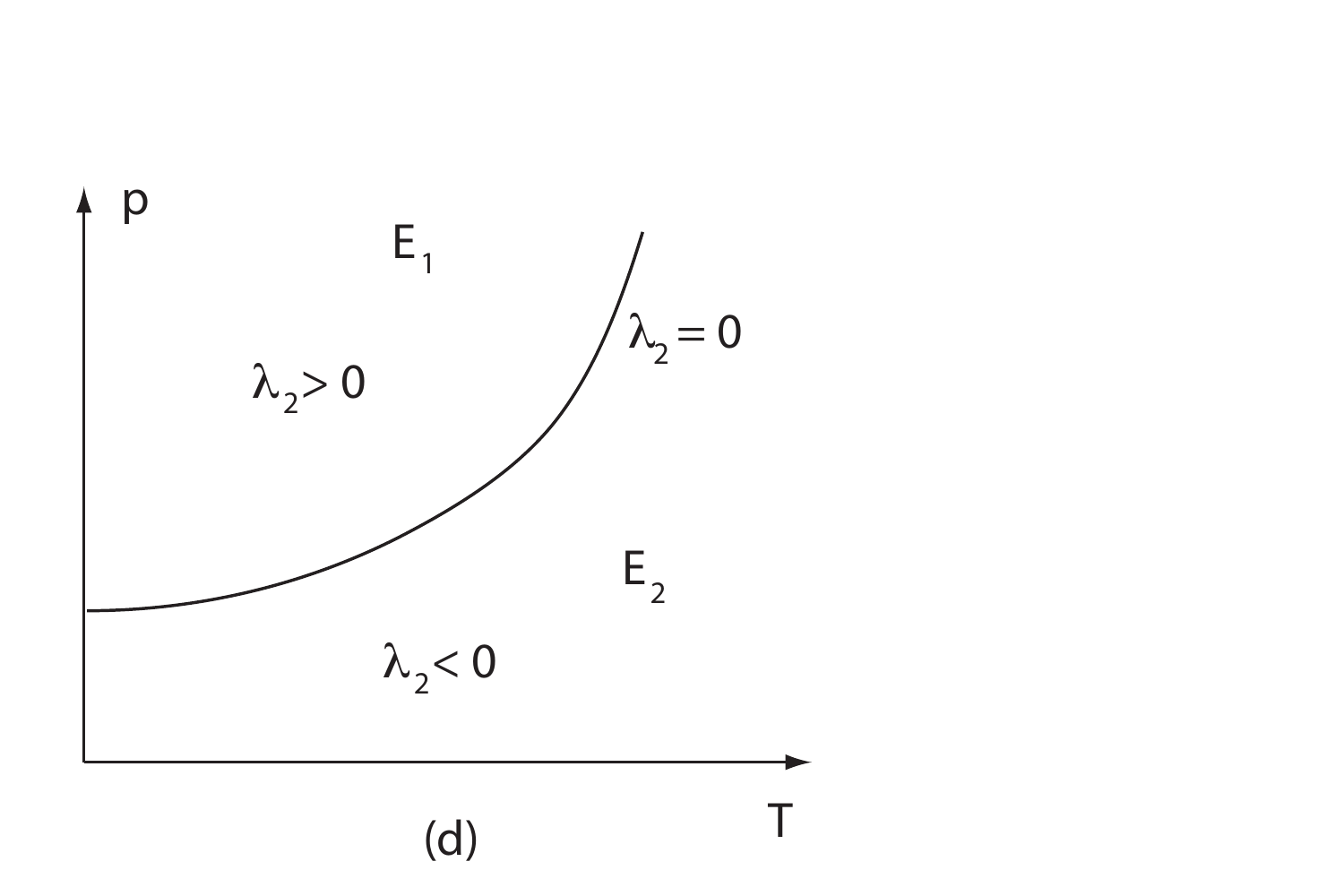}
  \caption{(a) The curve of $\gamma_1=0$, (b) the curve of
$\mu_1=0$, (c) the curve of $\lambda_1=0$, (d) the curve of
$\lambda_2=0$.}\la{f8.34}
 \end{figure}

\begin{SCfigure}[25][t]
  \centering
  \includegraphics[width=0.35\textwidth]{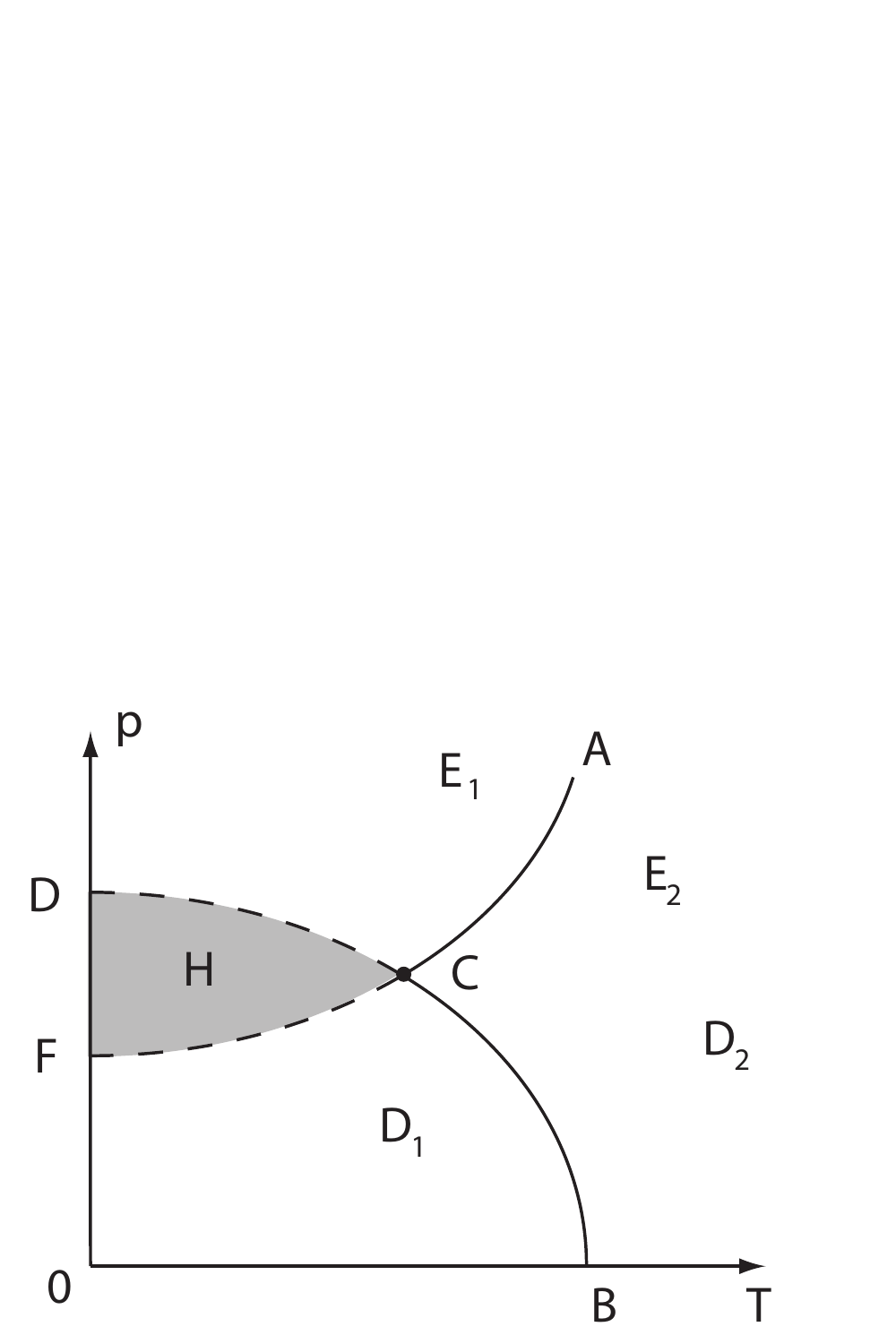}
  \caption{$ACF$ is the curve of $\lambda_2=0, BCD$ is
the curve of $\lambda_1=0$.}\la{f8.35}
 \end{SCfigure}

Let 
\begin{eqnarray*}
&&D_1=\{(T,p)\in \R^2_+\ |\ \lambda_1(T,p)>0\},\\
&&D_2=\{(T,p)\in \R^2_+\ |\ \lambda_1(T,p)<0\},\\
&&E_1=\{(T,p)\in \R^2_+\ |\ \lambda_2(T,p)>0\},\\
&&E_2=\{(T,p)\in \R^2_+\ |\ \lambda_2(T,p)<0\}.
\end{eqnarray*}
Let the curve of $\lambda_1(T,p)=0$ intersect with the curve of
$\lambda_2(T,p)=0$ at a point $C$; see Figure \ref{f8.35}. 

When $\lambda_1$ crosses the curve segment $CB$
to enter into $D_1$ from $D_2$ (see Figure \ref{f8.35}),  we have $\lambda_2<0$ and $\lambda_1$
satisfies
$$
\lambda_1(T,p)\left\{\begin{aligned}
&<0 &&\text{ if }\ (T,p)\in D_2,\\
&=0 &&\text{ if }\ (T,p)\in CB,\\
&>0 &&\text{ if }\ (T,p)\in D_1.
\end{aligned}
\right.
$$ 
In this case, by Theorem \ref{t5.1}, the equations
(\ref{8.207}) have a phase change which describes the transition
from liquid He-I to liquid He-II, and the second equation of
(\ref{8.207}) can be equivalently rewritten as
\begin{equation}
\frac{d\widetilde{\rho_n}}{dt}=(\lambda_2-2b_2\rho^*_n-3b_3(\rho^{*}_n)^2)\widetilde{\rho_n}-
(b_2+3b_3\rho^*_n)\widetilde{\rho}^2_n-b_3\widetilde{\rho}^3_n,\label{8.215}
\end{equation}
where $\widetilde{\rho}_n=\rho_n-\rho^*_n$,  $\rho^*_n$
satisfies the equation
$$\lambda_2\rho_n-b_2\rho^2_n-b_3\rho^3_n=b_1\rho_s,$$
and  $\rho_s>0$ is the transition solution of (\ref{8.207}). Since
$\rho^*_n$ is the density deviation of normal liquid, by
(\ref{8.199}) we have
$$\rho^*_n=\rho -\rho^0_n-\rho_s.$$
On the other hand, $\rho^0_n \ (\simeq\rho )$ represents the density
of liquid He I. Thus, $\rho^*_n\simeq -\rho_s<0$. From the
$PT$-phase diagram of $^4$He (Figure \ref{f8.33}), we see that
$\rho_{sol}\geq\rho_l$ for $^4$He near $T=0$, therefore by
(\ref{8.214}), $b_2\leq 0$. Hence we derive that
\begin{equation}
\lambda_2-2b_2\rho^*_n-3b_3(\rho^*_n)^2<0,\label{8.216}
\end{equation}
for $(T,p)$ in the region $D_1 \setminus H$, as shown in Figure \ref{f8.35}.

It follows from (\ref{8.215}) and (\ref{8.216}) that when $(T,p)$
is in the region $D_1 \setminus H$, the liquid $^4$He is in the superfluid
state.

When $\lambda_2$ crosses the curve segment $CA$, as shown in
Figure \ref{f8.35}, to enter into $E_1$ from $E_2$, then $\lambda_1<0$
and 
$$\lambda_2(T,p)
\left\{\begin{aligned}
& <0  && \text{ if }\ (T,p)\in E_2,\\
& =0  && \text{ if }\ (T,p)\in CA,\\
& >0  && \text{ if }\ (T,p)\in E_1.
\end{aligned}
\right.
$$ 
In this case, the equations (\ref{8.207}) characterize
the liquid-solid phase transition, and the first equation of
(\ref{8.207}) is equivalently expressed as
\begin{equation}
\frac{d\rho_s}{dt}=(\lambda_1-a_1\widetilde{\rho}_n)\rho_s-a_2\rho^2_s,\label{8.217}
\end{equation}
where $\widetilde{\rho}_n>0$ is the transition state of solid
$^4$He . By (\ref{8.212}) we have
\begin{equation}
\lambda_1(T,p)-a_1\widetilde{\rho}_n<0,\label{8.218}
\end{equation}
for any $(T,p)$ in the region $E_1 \setminus H$, as shown in Figure \ref{f8.35}.

From (\ref{8.217}) and (\ref{8.218}) we can derive the conclusion
that as $(T,p)$ in $E_1 \setminus H, \rho_s=0$ is stable, i.e., $^4$He is in
the solid state.

However, the shadowed region $H$  in Figure~\ref{f8.35}  is an unstable domain for the
solid and liquid He II states, where any  of these two
phases may appear depending on the random fluctuations. Thus, from
the discussion above, we can derive the theoretical $PT$-phase
diagram given by Figure \ref{f8.36}, based on equations
(\ref{8.207}). In  comparison  with the experimental $PT$-phase diagram
(Figure \ref{f8.33}), a sight difference in Figure \ref{f8.36} is that there
exists an unstable region $H$,  where the solid phase and the
He II phase are possible to occur. This unstable region in Figure
\ref{f8.36} corresponding to the coexistence curve $CE$ in Figure \ref{f8.33} is
a theoretical prediction, which need to be verified by
experiments.
\begin{SCfigure}[25][t]
  \centering
  \includegraphics[width=0.4\textwidth]{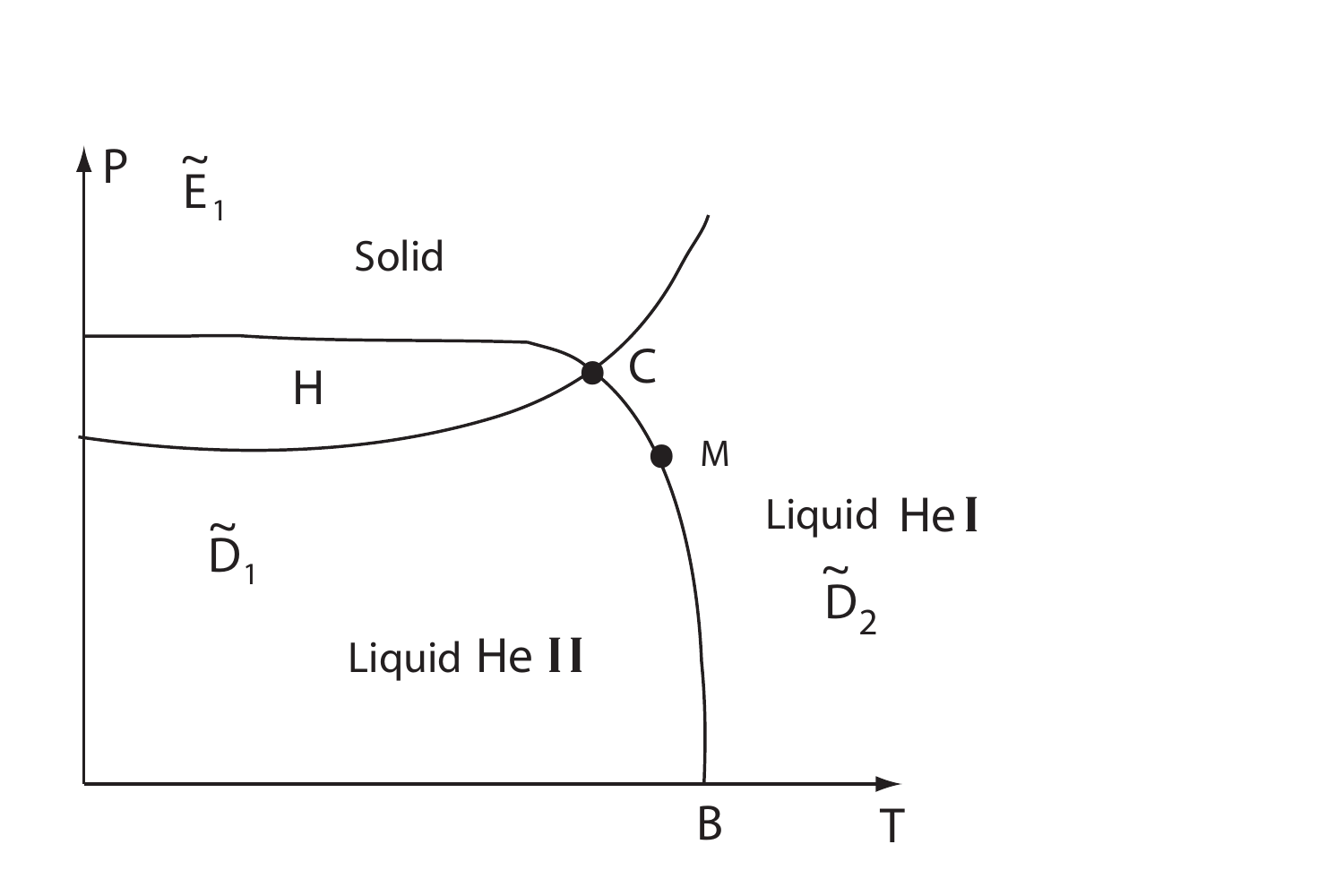}
  \caption{Theoretical $PT$-phase diagram:  $H$ is an
unstable region where both solid and liquid He II states appear randomly depending on fluctuations. The point $M$ is a point where the transitions,  between superfluid state (liquid He II) and the normal fluid state (liquid He I), changes from Type II to Type-I; namely, the transition crossing $CM$ is Type-II and the transition crossing $MB$ is Type-I.}\la{f8.36}
 \end{SCfigure}

\subsection{States in unstable region}

We consider the dynamical properties of transitions for
(\ref{8.207}) in the unstable region. It is clear that at point
$C=(T_C,p_C)$,
\begin{equation}
\lambda_1(T_C,p_C)=0,\ \ \ \ \lambda_2(T_C,p_C)=0,\label{8.219}
\end{equation}
and the unstable region $H$ satisfies that
$$H=\{(T,p)\in \R^2_+\ |\ \lambda_1(T,p)>0,\ \lambda_2(T,p)>0\}.$$

To study the structure of flows of (\ref{8.207}) for $(T,p)\in H$
it is necessary to consider the equations (\ref{8.207}) at the  point
$C=(T_C,p_C)$, and by (\ref{8.219}) which are given by
\begin{equation}
\left.
\begin{aligned}
&\frac{d\rho_s}{dt}=-a_1\rho_n\rho_s-a_2\rho^2_s,\\
&\frac{d\rho_n}{dt}=-b_1\rho_s-b_2\rho^2_n-b_3\rho^3_n.
\end{aligned}
\right.\label{8.220}
\end{equation}

Due to (\ref{8.214}) and $\rho_{sol}>\rho_l$, we have
\begin{equation}
b_2<0\ \ \ \ \text{for}\ \ \ \ (T,p)\subset H.\label{8.221}
\end{equation}
Under the condition (\ref{8.221}), equations (\ref{8.220}) have
the following two steady state solutions:
\begin{eqnarray*}
&&Z_1=(\rho_s,\rho_n)=(0,|b_2|/b_3),\\
&&Z_2=(\rho_s,\rho_n)=\left(\frac{a_1}{a_2}\alpha ,-\alpha
\right),\\
&&\alpha =\frac{|b_2|}{2b_3}(\sqrt{1+4a_1b_1b_3/a_2|b_2|^2}-1).
\end{eqnarray*}
By direct computation, we can prove that the eigenvalues of the
Jacobian matrices of (\ref{8.220}) at $Z_1$ and $Z_2$ are
negative. Hence, $Z_1$ and $Z_2$ are stable equilibrium points of
(\ref{8.220}). Physically, $Z_1$ stands for solid state, and $Z_2$
for superfluid state. The topological structure of (\ref{8.220})
is schematically illustrated by Figure \ref{f8.37}(a), the two regions
$R_1$ and $R_2$ divided by curve $AO$ in Figure \ref{f8.37}(b) are the
basins of attraction of $Z_1$ and $Z_2$ respectively.

We note that in $H, \lambda_1$ and $\lambda_2$ is small, i.e.,
$$0<\lambda_1(T,p),\ \ \ \ \lambda_2(T,p)\ll 1,\ \ \ \ \text{for}\
(T,p)\in H,$$ and (\ref{8.207}) can be consider as a perturbed
system of (\ref{8.220}).

Thus, for $(T,p)\in H$ the system (\ref{8.207}) have four steady
state solutions $\widetilde{Z}_i=\widetilde{Z}(T,p)$  $(1\leq i\leq
4)$ such that
$$\lim_{(T,p)\to (T_C,p_C)}   ( \widetilde{Z}_1(T,p),   \widetilde{Z}_2(T,p), 
\widetilde{Z}_3(T,p), \widetilde{Z}_4(T,p))=(
 Z_1,  Z_2, 0,  0),$$ 
 and $\widetilde{Z}_1$ and
$\widetilde{Z}_2$ are stable, representing solid state and liquid
He-II state respectively, $\widetilde{Z}_3$ and $\widetilde{Z}_4$
are two saddle points. The topological structure of (\ref{8.207})
for $(T,p)\in H$ is schematically shown in Figure \ref{f8.37}(c), and the
basins of attraction of $\widetilde{Z}_1$ and $\widetilde{Z}_2$
are $\widetilde{R}_1$ and $\widetilde{R}_2$ as illustrated by
Figure \ref{f8.37}(d).
\begin{figure}[hbt]
  \centering
  \includegraphics[width=0.45\textwidth]{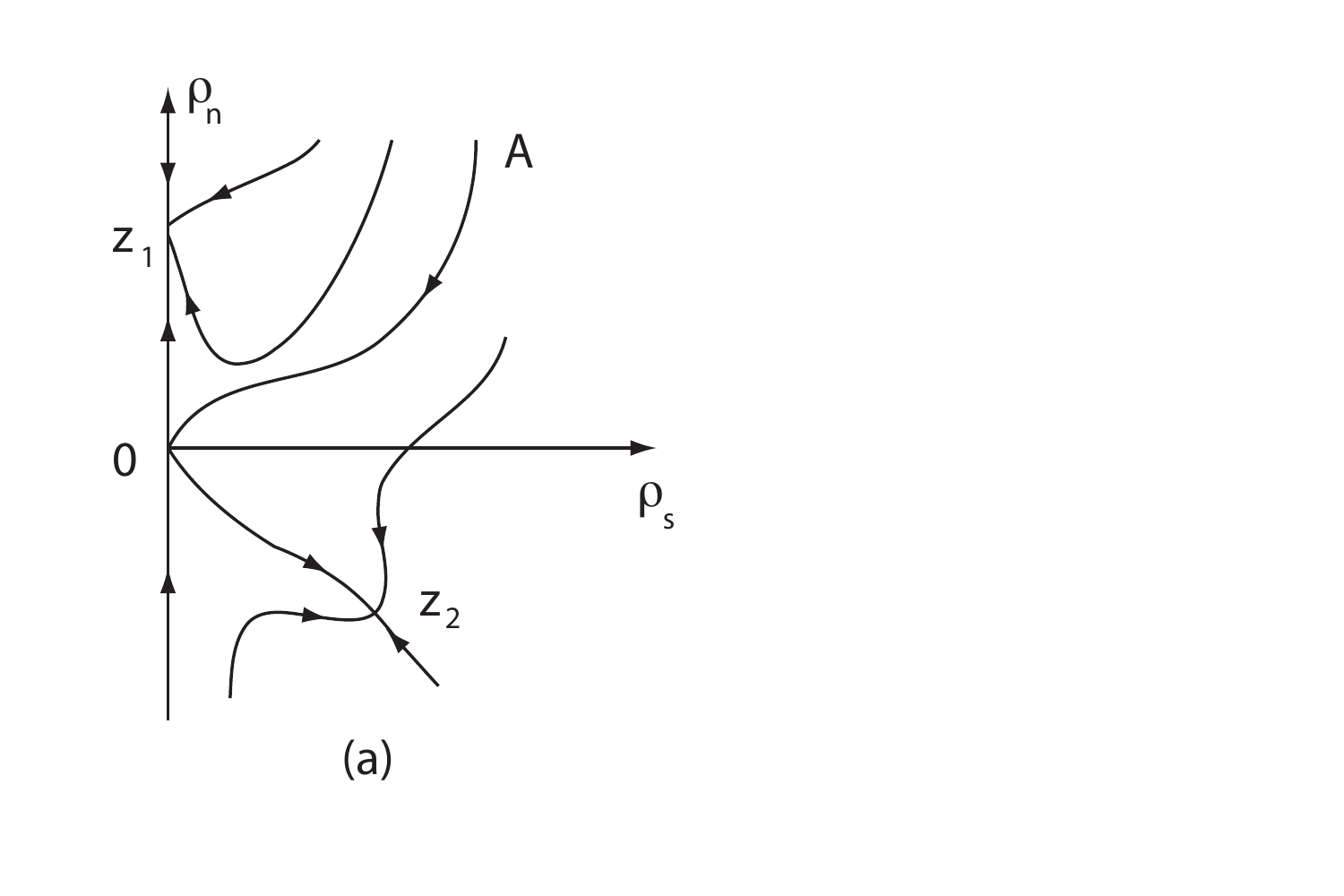}
  \includegraphics[width=0.3\textwidth]{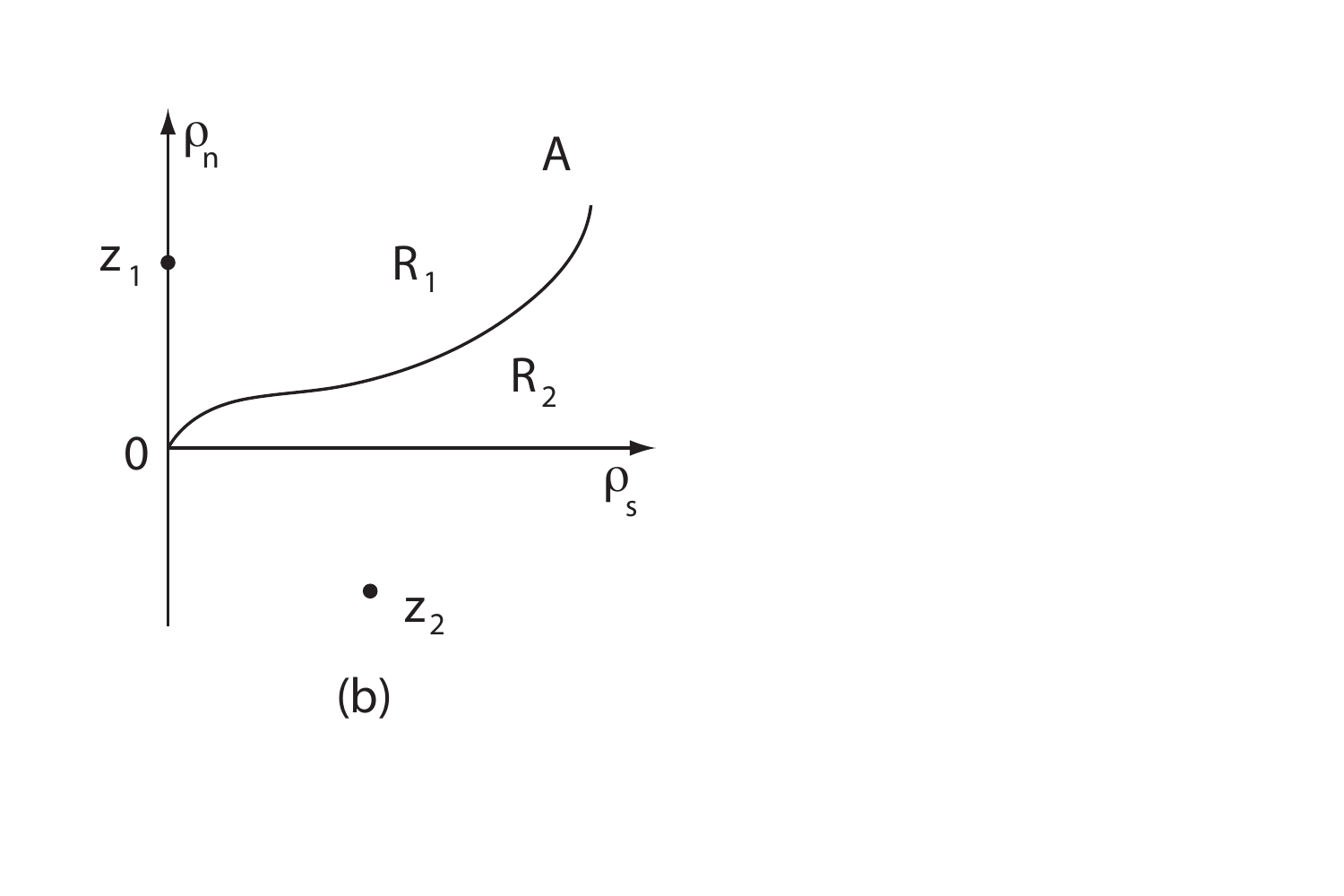}  \\
  \includegraphics[width=0.3\textwidth]{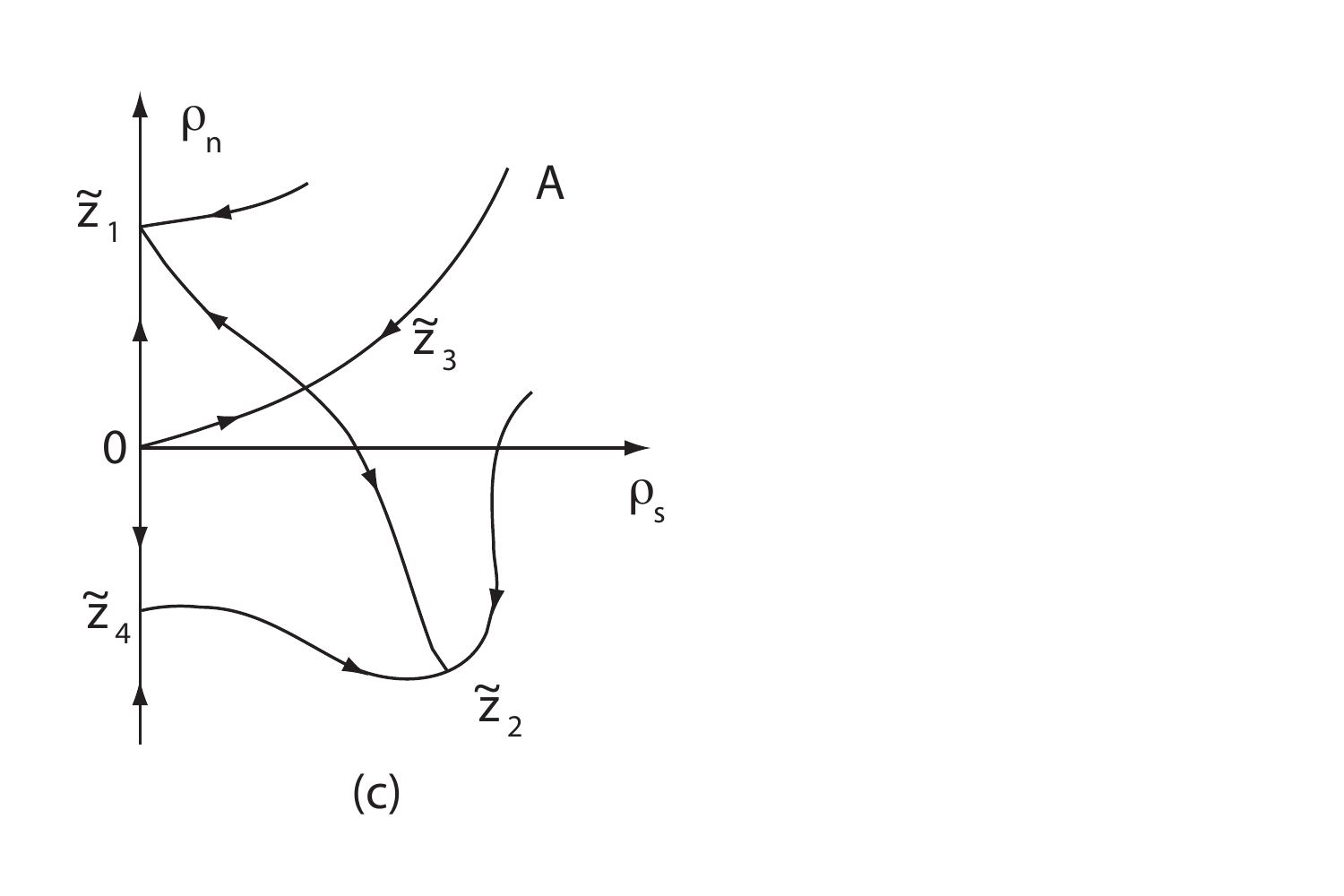}\qquad \quad
  \includegraphics[width=0.28\textwidth]{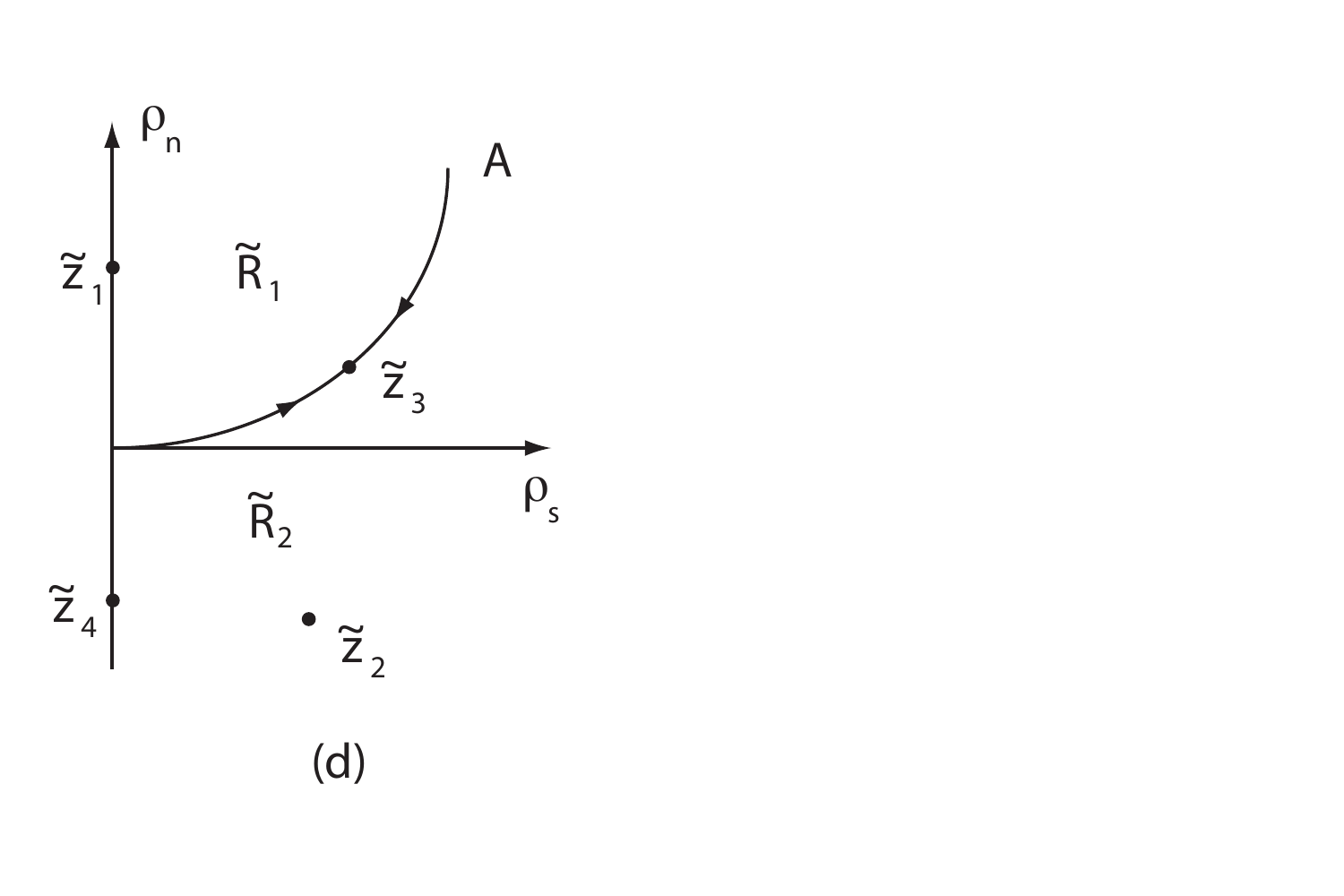}
  \caption{}\la{f8.37}
 \end{figure}

In summary, with the above analysis and the dynamic transition theory, we arrive at the following theorem:

\bt\la{t8.14-1}
There exist four regions
$\widetilde{E}_1,\widetilde{D}_1$ and $H$ in  the  $PT$-plane (see
Figure \ref{f8.36}), which are defined by
\begin{align*}
&\widetilde{E}_1=\{(T,p)\in \R^2_+\ |\ \lambda_1(T,p)<0,\
\lambda_2(T,p)>0\},\\
&\widetilde{D}_1=\{(T,p)\in \R^2_+\ |\ \lambda_1(T,p)>0,\
\lambda_2(T,p)<0\},\\
& \tilde D_2 = \{(T,p)\in \R^2_+\ |\ \lambda_1(T,p) < 0,\
\lambda_2(T,p)<0\},\\
&H=\{(T,p)\in \R^2_+\ |\ \lambda_1(T,p)>0,\ \lambda_2(T,p)>0\},
\end{align*}
such that the following conclusions hold true:

\begin{itemize}

\item[(1)] If $(T,p)\in\widetilde{E}_1$,  the phase of $^4$He is
in solid state.

\item[(2)] If  $(T,p)\in\widetilde{D}_1$,  the phase is in
superfluid state.

\item[(3)] If $(T, p) \in \widetilde{D}_2$, the phase is in the normal fluid state.

\item[(4)] If  $(T,p)\in H$, there are two regions
$\widetilde{R}_1$ and $\widetilde{R}_2$ in the state space
$(\rho_s,\rho_n)$  such that, under a fluctuation which is described by the
initial value $(x_0,y_0)$ in (\ref{8.207}), if
$(x_0,y_0)\in\widetilde{R}_1$ then the phase is in solid state,
and if $(x_0,y_0)\in\widetilde{R}_2$ then it is in superfluid
state.
\end{itemize}
\et

 In general, the observed superfluid transitions of $^4$He are of the
 second order, i.e., Type-I (continuous) with  the dynamic classification. 
 But, from (\ref{8.207}) we can
 prove the following theorem, which shows that for a higher pressure 
 the superfluid transitions may be of zeroth 
 order, i.e., the  Type-II (jump) transition with the dynamic classification scheme.
 
 \bt\la{t8.14}
  Let $(T_0,p_0)$ satisfy that
 $\lambda_1(T_0,\rho_0)=0, \lambda_2(T_0,p_0)<0$. Then,
 (\ref{8.207}) have a superfluid transition at $(T_0,p_0)$  from $D_2$  to $D_1$. In particular, the following assertions hold true:
 
 \begin{itemize}
 
 \item[(1)] Let  \begin{equation}
 A=\frac{a_1b_1}{|\lambda_2|}-a_2\ \ \ \ \text{at}\
 (T,p)=(T_0,p_0).\label{8.221-1}
 \end{equation}
 Then if $A<0$ the superfluid transition is Type-I, and if $A>0$
 the superfluid transition is Type-II.
 
 \item[(2)]  The equation 
 $$A=\frac{a_1b_1}{|\lambda_2|}-a_2=0$$
 determines a point $M$ on the $CB$ coexistence curve in Figure~\ref{f8.36}, where the transition changes type from Type-II to Type-I.
 \end{itemize}
 
 \et
 
\bp 

{\sc Step 1.} 
 By  the  assumption, it is clear that
 $$\lambda_1(T,p)=
 \left\{\begin{aligned}
&  <0 &&\text{ if }\ (T,p)\in D_2,\\
& =0 &&\text{ if }\ (T,p)=(T_0,p_0),\\
& >0 &&\text{ if }\ (T,p)\in D_1,
 \end{aligned}
 \right.$$
 where $D_1,D_2$ are as in Figure \ref{f8.34}(c). Hence, by Theorem \ref{t5.1},
 (\ref{8.207}) have a transition at $(T_0,p_0)$. By
 $\lambda_2(T_0,p_0)<0$, the first eigenvalue of (\ref{8.207}) is
 simple, and its eigenvector is given by
 $e=(e_s,e_n)=(1,0).$

 The reduced equation of (\ref{8.207}) on the center manifold reads
 $$\frac{dx}{dt}=\lambda_1x-a_1xh(x)-a_2x^2,\ \ \ \ x>0,$$
 where $h(x)$ is the center manifold function. By  (\ref{8.207}), $h$ 
  can be expressed as
 $$h(x)=\frac{b_1}{\lambda_2}x+o(x^2),\ \ \ \ x>0.$$
 Thus the reduced equation of (\ref{8.207}) is given by
 \begin{equation}
 \frac{dx}{dt}=\lambda_1x+Ax^2+o(x^2) \qquad \text{ for } x > 0, \label{8.222}
 \end{equation}
 where $A$ is as in (\ref{8.221-1}). The theorem follows from
 (\ref{8.222}) and  Theorem \ref{t5.3}. 

\medskip 
{\sc Step 2.} By the nondimensional form, we see that
 $$a_1b_1=\frac{1}{2}\gamma^2_3|\psi_0|^2\tau^2,$$
 where $\gamma_3$ is the coupled coefficient of $\rho_n$ and $\rho_s$.
 Physically, $\gamma_3$ is small in comparison with $\gamma_2$; namely
 $$0<a_1b_1\ll a_2.$$
 On the other hand, we know that
 $$\lambda_2(T_0,p_0)\rightarrow 0\ \ \ \ \text{as}\ \ \ \
 (T_0,p_0)\rightarrow (T_C,p_C),$$
 where $C=(T_C,p_C)$ is as in (\ref{8.219}). Therefore we deduce
 that there exists a pressure $p^*(b<p^*<p_C)$ such that
 $$A=\frac{a_1b_1}{|\lambda_2(T_0,p_0)|}-a_2
 \left\{\begin{aligned}
&  <0 && \text{ if }\ 0\leq p_0<p^*,\\
 &  >0 && \text{ if } \ p^*<p_0<p_C.
 \end{aligned}
 \right.$$
 Thus,  when  the transition pressure
 $p_0$ is below some value $p_0<p^*$ the superfluid transition is of 
 the second order, i.e., is continuous with the dynamic classification scheme, and when $p_0<p^*$ the
 superfluid transition is the first order, i.e., is jump in the dynamic classification scheme.
 The proof is complete.
 \ep

\section{Physical Conclusions and Predictions}
As we know, the classical phase transition is illustrated by Figure~\ref{f8.33}. In this phase diagram, the coexistence curve $CE$ separates the solid state and the superfluid state (liquid He II), and the curve $CB$ is the coexistence curve between the superfluid state (liquid He II) and the normal fluid state (liquid He I), and the critical point is the 
triple-point. It is considered by the classical theory that the transition crossing $CB$ is second order in the Ehrenfest sense (Type-I with the dynamic classification scheme). 

However, in the phase transition diagram Figure~\ref{f8.36} derived based on Theorems~\ref{t8.14-1}  and \ref{t8.14}, 
there is a unstable region $H$, where both solid and liquid He II states appear randomly depending on fluctuations. The point $M$ is a {\it switch point} where the transitions,  between superfluid state (liquid He II) and the normal fluid state (liquid He I), changes from first order (Type II with the dynamic classification scheme)  to second order (Type-I); namely, the transition crossing $CM$ is second order (Type-II) and the transition crossing $MB$ is first  order (Type-I).

\medskip

In summary, the results in this article predict the existence of the unstable region $H$  and the existence of the switch point $M$. It is hoped these predictions can be verified by experiments.
\bibliographystyle{siam}
\def\cprime{$'$}

\end{document}